\documentclass[12pt,preprint,a4paper]{aastex}
\usepackage{times}
\usepackage{graphics,epsf}
\usepackage{amsmath}                
\usepackage{amsfonts}               
\usepackage{amssymb}                
\usepackage{epsfig}                 
\usepackage{graphicx}
\usepackage{url}
\usepackage{tabularx}
\usepackage{booktabs}

\def \s{~\rm{s}}
\def \ms{~\rm{ms}}

\def \km{~\rm{km}}

\def \g{~\rm{g}}

\def \erg{~\rm{erg}}
\def \foe{~\rm{foe}}

\def \MeV{~\rm{MeV}}

\begin{document}

\title{A call for a paradigm shift from neutrino-driven to
jet-driven core-collapse supernova mechanisms}

\author{Oded Papish\altaffilmark{1} Jason Nordhaus\altaffilmark{2,3}and Noam Soker\altaffilmark{1}}

\altaffiltext{1}{Department of Physics, Technion -- Israel Institute of Technology, Haifa
32000, Israel; papish@physics.technion.ac.il; soker@physics.technion.ac.il}
\altaffiltext{2}{Department of Science and Mathematics, National Technical Institute for the Deaf, Rochester Institute of Technology, Rochester, NY, USA; nordhaus@astro.rit.edu}
\altaffiltext{3}{Center for Computational Relativity and Gravitation, Department of Mathematics, Rochester Institute of Technology, Rochester, NY, USA}
\begin{abstract}
{Three-dimensional (3D) simulations in recent years have shown
severe difficulties producing $10^{51} \erg$ explosions of massive
stars with neutrino based mechanisms while on the other hand
demonstrated the large potential of mechanical effects, such as
winds and jets in driving explosions. In this paper we study the
typical time-scale and energy for accelerating gas by neutrinos in
core-collapse supernovae (CCSNe) and find that under the most
extremely favorable (and probably unrealistic) conditions, the
energy of the ejected mass can reach at most $5 \times 10^{50}$
erg. More typical conditions yield explosion energies an
order-of-magnitude below the observed $10^{51}$ erg explosions. On
the other hand, non-spherical effects with directional outflows
hold promise to reach the desired explosion energy and beyond.
Such directional outflows, which in some simulations are produced
by numerical effects of 2D grids, can be attained by angular
momentum and jet launching. Our results therefore call for a
paradigm shift from neutrino-based explosions to jet-driven
explosions for CCSNe.}
\end{abstract}

\section{INTRODUCTION}
\label{sec:zintro}

Eighty years after \cite{Badde1934} first suggested that
supernovae (SNe) are powered by stars collapsing into neutron
stars (NS), the processes by which part of this gravitational
energy is channelled to explosion remains controversial.
\cite{Wilson1985} and \cite{bethe1985} refined the neutrino
mechanism \citep{Colgate1966} into the delayed-neutrino mechanism,
whereby neutrinos emitted within a period of $\sim$$1 \s$ after
the bounce of the collapsed core heat material in the gain region
($r \approx 100-200 \km$).  This subsequent neutrino-heating was
thought to revive the stalled shock thereby exploding the star and
producing a canonical core-collapse supernova (CCSN) with an
observed energy of $E_{\rm exp} \ga 1 \foe$, where $1 \foe \equiv
10^{51} \erg$.

In the last three decades, sophisticated multidimensional
simulations with increasing capabilities were used to study the
delayed-neutrino mechanism (e.g.,
\citealt{bethe1985,Burrows1985,Burrows1995,Fryer2002,Buras2003,Ott2008,Marek2009,Nordhaus2010,Brandt2011,Hanke2012,Kuroda2012,
Hanke2012,Mueller2012,Bruenn2013,MuellerJanka2014,
Mezzacappaetal2014, Bruennetal2015}). The outcome of such
numerical experiments varied widely with many failing to revive
the stalled shock while others produced tepid explosions with
energies less than 1~$\foe$. Historically, in spherically
symmetric calculations (1D), the vast majority of progenitors can
not even explode
\citep{Burrows1995,Rampp2000,Mezzacappa2001,Liebend2005}. The
exception being the 8.8-$M_{\odot}$ progenitor of
\citet{Nomoto1988} which resulted in a $\sim$3$\times10^{49}$ erg
neutrino-driven-wind explosion due to the rarified stellar
envelope \citep{Kitaura2006}. Extension to axisymmetric
calculations (2D) yielded similar outcomes over their 1D
counterparts despite the inclusion of instabilities such as
neutrino-driven convection and the standing-accretion-shock
instability (SASI)
\citep{Burrows1995,Janka1996,Buras2006a,Buras2006b,Ott2008,Marek2009}.

It should be noted that while many of the current numerical
experiments incorporate multi-dimensional hydrodynamics,
performing 3D radiation is currently prohibitive computationally
\citep{Zhang2013}.  Many groups utilize multi-group-flux-limited
diffusion (MGFLD) in the 1D ``ray-by-ray" transport approximation.
This is a reasonable approach to core-collapse simulations both
because of the limitation of current computational resources and
because the results for multi-angle transport are similar to those
for MGFLD except in the cases of extremely rapid rotation
\citep{Ott2008}.  Thus, it's unlikely that future simulations that
incorporate 3D transport will yield fundamental differences over
current state-of-the-art calculations in terms of the viability of
neutrino mechanism.

Recently, a number of groups have published 3D core-collapse
simulations with differing computational approaches and various
levels of sophistication \citep{Nordhaus2010,Janka2013,
Couch2013,Dolence2013,Takiwakietal2013, Dolenceetal2014,Hanke2012,
Hanke2013, couch2013arXiv, Mezzacappaetal2014}. Some groups find
that the extra-degree of freedom available in 3D simulations makes
it easier to achieve shock revival over their axisymmetric
counterparts \citep{Nordhaus2010,Dolence2013}.  On the other hand,
several groups have found the opposite; namely that explosions are
harder to achieve in 3D than 2D \citep{Janka2013,
Couch2013,Takiwakietal2013, Hanke2012, Hanke2013, couch2013arXiv}.
If that's the case, then it may well be that the delayed-neutrino
mechanism categorically fails and alternative mechanisms should be
investigated.

In one recent case, axisymmetric calculations of 12-, 15-, 20-,
and 25-$M_\odot$ progenitors successfully revived the shock with
explosion energy estimates of $\sim$ $0.3-0.9 \foe$
\citep{Bruenn2013, Mezzacappaetal2014, Bruennetal2015}. {Their
energy is supplied primarily by an enthalpy flux. This is actually
a wind, mainly along the imposed symmetry axis, i.e., a collimated
wind. This wind is driven by the inflowing (accreted) gas.}  Winds
were suggested to power CCSN in the past (e.g.,
\citealt{Burrows1993, Burrows1995}), but were found to have
limited contribution to the explosion for a more massive than $8.8
M_\odot$ starts.

Many CCSNe, e.g., some recent Type Ic SNe
(\citealt{Roy2013,Takaki2013}) explode with kinetic energy of $\ga
10 \foe$.  Neutrino based mechanisms cannot account for such
energies even under favorable conditions. For example,
\cite{Ugliano2012} performed a set of simulations where the energy
was artificially scaled to that of SN 1987A, and found that even
if neutrino explosions do work for some CCSNe, no explosions with
kinetic energy of $>$ $2\foe$ are achieved. {{{ This scaling was
achieved by artificially setting the inner boundary luminosity to
obtain an explosion with an energy equal to that of SN 1987A. }}}
The delayed-neutrino mechanism must be shown to produce robust
explosions with canonical supernova energies for a range of
progenitors if it is to continue to be a contender in
core-collapse theory. Despite decades of effort with the most
sophisticated physics to date, no current simulation has produced
a successful $10^{51}$ erg supernova. It is this fact that leads
us to argue that the delayed-neutrino mechanism has a generic
character that prevents it from exploding the star with an
observed energy of $1\foe$.


The delicate and problematic nature of neutrino-driven mechanisms
were already revealed with 1D simulations, such that even the most
sophisticated neutrino transport calculations were unable to
explode stars for progenitor masses $\gtrsim$ $12 M_\odot$ (e.g.,
\citealt{Liebendorferetal}). Multidimensional effects were then
seen as necessary for triggering an explosion. The most common
multi-dimensional processes that have been studied as a rescue for
the delayed-neutrino mechanism were neutrino-driven convection
(e.g., \citealt{Burrows1995}) and hydrodynamic instabilities, such
as the SASI \citep{BlondinMezzacappa2003}. These axisymmetic (2D)
simulations have shown mixed and contradicting results. Most do
not get an `explosion' at all, while others obtain explosions with
very little energy, i.e., $\ll 1 \foe$ (e.g.,
\citealt{Suwa2013,Suwaetal2010}). In most of these cases where an
`explosion' is claimed, it is actually only shock revival and not
a typical explosion, as the energy is much too low to explain most
observed CCSNe.

In the past few years, the regime of 3D flow structures have been
explored in more detail (e.~g. \citealt{Nordhaus2010}). The
simulations have not reached any consensus on the outcome. While
some show that it is easier to revive the shock in 3D than in 2D
{{{{(e.g., \citealt{Nordhaus2010,Dolence2013}),} others showed the
opposite {{{{(e.g.,
\citealt{Hanke2013,Couch2013ApJ,couch2013arXiv,Takiwakietal2013}).}
Even in 3D simulations that successfully revive the shock, the
energy is significantly lower than $1 \foe$. {{{{Recently,
turbulence from convective burning in the Si/O shell were shown to
aid shock revival (\citealt{CouchOtt2014,MuellerJanka2014b}).} }}}

{{{ A recent demonstration of outcome sensitivity on initial
setting are the two 3D studies by \cite{Nakamuraetal2014} and
\cite{mostaetal2014}. \cite{Nakamuraetal2014} find an explosion
energy of $\sim 1 \foe$ for a case with a rapid core rotation. For
a rotation velocity of $0.2$ times that rapid rotation, the
explosion energy was only $\sim 0.1 \foe$. They did not include
magnetic fields. \cite{mostaetal2014} included very strong
magnetic fields in the pre-collapse core as well as a very rapid
rotation, about twice as large as the rapid rotation case of
\cite{Nakamuraetal2014}. \cite{mostaetal2014} obtained jets but
did not manage to revive the stalled shock and did not obtain any
explosion. }}}

The structure of this paper is as follows.  In section
\ref{sec:Time}, we expand upon the argument presented in
\cite{Papish2012a} that the delayed-neutrino mechanism cannot
achieve canonical supernova energies. We consider the limitation
of the delayed-neutrino mechanism from another perspective in
section  \ref{sec:Energy}. In section \ref{sec:shock} we discuss
the role of progenitor perturbations and why contradicting results
are common among the groups simulating neutrino-based mechanisms,
and in section \ref{sec:recombination} we discuss the energy
available from recombination of free nucleons. A discussion of
{the collimated-wind obtained by \cite{Bruennetal2015} } and our
summary are in section \ref{sec:summary}.

\section{TIME-SCALE CONSIDERATIONS}
\label{sec:Time}

We start with simple time-scale considerations during the revival of the shock in a spherically symmetric outflow.
The ``gain region" of the delayed neutrino mechanism, i.e. where neutrino heating outweighs neutrino cooling,
typically occurs in the region $r \simeq 100-400 \km$ \citep{Janka2001}.

{{{{  For an explosion to be initiated the advection timescale
$\tau_{\rm adv}$ should be larger than the heating timescale
$\tau_{\rm heat}$. This advection timescale is the time needed for
material to cross the gain region during accretion. Most
core-collapse simulations fail when this condition is not
fulfilled. When this condition is met the internal energy can
increase until there is enough energy to unbind the material and
an explosion is initiated. At this point the total energy of the
gas in the gain region is very close to zero. From this time the
net heating adds up to the positive explosion energy. After the
gas reaches large radii, $\ga 1000 \km$, heating becomes
inefficient. It is true that some gas expands at a lower velocity
and it is closer to the center. However, density decreases and so
does the neutrino {{{{{ optical depth that decreases below its
initial value, such that neutrino heating becomes even less
efficient. Material near the neutrinosphere has, by definition,
a large optical depth. It can in principle absorb energy  and
expand. But this process is a neutrino-driven wind, which is not
part of the delayed-neutrino mechanism, and was found
to have limited contribution to the explosion (e.g.,
\citealt{Burrows1993, Burrows1995}). }}}}}
 The time from the start of acceleration to the end of
efficient heating is marked $t_{\rm esc}$. From simulations
$t_{\rm est} \simeq 50 \ms$ \citep{Bruenn2013, Bruennetal2015,
Marek2009}. In section \ref{sec:Energy} we find a similar time
from a simple analytical estimate.  }}}}

{{{{ In figure 2 of \cite{Bruenn2013} the shock is starting to
expand and an explosion is initiated at time $t\simeq 200 \ms$. At
this time the total positive energy is close to zero (figure 4 in
\citealt{Bruenn2013}). At that time the shock is at a distance of
$r_{\rm s} \simeq 400 \km$. This shows that during the time the
shock moves from $200 \km$ to $400 \km$ the total energy increased
from a negative value to about zero. We take the time of zero
energy to be the starting point of postive energy accumulation,
and use it to estimate the explosion energy. In the simulations of
\cite{Bruenn2013} at time $t=300 \ms$ the shock is already at a
distance of $r_{\rm s} \simeq 1000-1500 \km$. Some material is
closer to the center, but its density is lower than at earlier
times, opacity is lower, and heating is inefficient. We note again
the long duration of energy increase in the work of
\cite{Bruenn2013, Bruennetal2015} and \cite{ Mezzacappaetal2014}, where
energy increases linearly with time for over a second, a time when
the shock is already at a distance of $r_{\rm s} \simeq 10,000
\km$. This linear growth of the energy can be explained by a
strong neutrino driven wind from the proto-neutron star. In the
new 3D case presented by \cite{Mezzacappaetal2014} the shock
radius position is similar to their results of 1D simulations
where no explosion have been obtained. }}}}

{{{{  A similar dynamic can be seen in figure 4 of the 2D
simulation of \cite{Marek2009}, where at time $t=524 \ms$ the
shock is at a radius of $r_{\rm s} \simeq 200 \km$. The shock
moves outward to $400 \km$ at $t=610 \ms$, but then at time $t=650
\ms$ the shock radius decrease back to $200 \km$. This shows that
at that time the energy is about zero and is not positive. The
acceleration time can be inferred from figure 6 where the average
shock moves from $400 \km$ to $700 \km$ during $\sim 50 \ms$. In
each direction the acceleration time lasts for $\sim 50 \ms$.
However, as the acceleration occurs at different times at
different directions, the behavior of the average shock radius
gives the impression that the acceleration phase is longer than
$50 \ms$.  }}}}

{{{{{For a neutrinoshpere at $r_{\nu} \simeq 50 \km$ (e.g.,
\citealt{couch2013arXiv}) }}}}} the neutrino ``optical depth''
from $r$ to infinity is given by
\begin{equation}
\tau_\nu \simeq 0.1(r/100 \km)^{-3}
\label{eq:depth}
\end{equation}
 \citep{Janka2001}, where the typical electron neutrino luminosity is $L_{\nu} \simeq L_{\bar \nu} \simeq 5 \times 10^{52} \erg \s^{-1}$ (e.g., \citealt{Mueller2012}).
Over all, if the interaction occurs near a radius $r$ in the gain region, the energy that can be acquired by the expanding gas is
\begin{equation}
E_{\rm shell} \simeq t_{\rm esc} \tau L_{\nu}  \simeq
0.25  \left( \frac{t_{\rm esc}}{50 \ms} \right) \left( \frac{L_{\nu}}{5 \times 10^{52} \erg \s^{-1}} \right)
\left( \frac{r}{100 \km} \right)^{-3}
\foe.
\label{eq:Eshell}
\end{equation}

Using a more typical radius of $\sim$ $200 \km $ for the
acceleration region, reduces the total energy to $0.03 \foe$.
Non-spherical flows that allow some simultaneous inflow-outflow
structure, might under favorable conditions be expected to
increase the energy by a factor of few to $\sim$ $0.1-0.3 \foe$.
This is consistent with numerical simulation results of the
delayed neutrino mechanism summarized in section \ref{sec:zintro}.
It is interesting to note that \cite{bethe1985} found an explosion
energy limit of $0.4 \foe$. This was based on their simulations
and not on any physical reason why the neutrino mechanism fails.

\section{ENERGY CONSIDERATIONS}
\label{sec:Energy}

We examine the situation by considering in more detail the acceleration from the delayed-neutrino mechanism.
Consider a mass $M_a$ that is accelerated and ejected by absorbing a fraction $f$ of the neutrino energy.
The mass starts at radius $r_0$ with zero energy. Namely the sum of internal and gravitational energy is zero.
This is an optimistic assumption, as the internal energy itself also needs to be supplied by neutrinos.
Neutrino losses can be absorbed into the parameter $f$.
After an acceleration time $t$ the energy of the mass is $ f L_ \nu t$ and its velocity is
\begin{equation}
v= \frac {dr}{dt} \simeq \left( \frac{2fL_\nu t}{M_a} \right)^{1/2}.
\label{eq:v}
\end{equation}
{{{{ Here we assume that most of the energy is transferred to
kinetic energy. Initially, more energy can be stored as thermal
energy. However, not much thermal energy can be stored after the
gas energy becomes positive, as it starts to accelerate outward
and thermal energy is converted to kinetic energy on a dynamical
time scale. The thermal energy acts to overcome gravity. We
calculate here the extra energy that goes to gas outward motion.
}}}}

Let the acceleration be effective to radius $r_a$ at time $t_a$. Integrating over time gives
\begin{equation}
r_a - r_ 0 \simeq  \frac{2}{3} \left( \frac{2fL_\nu}{M_a} \right)^{1/2} t_a^{3/2},
\label{eq:rer0}
\end{equation}
or
\begin{eqnarray}
& t_a \simeq
\left( \frac{9}{8} \right)^{1/3} \left(r_a - r_ 0 \right)^{2/3}  \left( \frac{M_a}{fL_\nu}\right)^{1/3}
\nonumber \\
& = 0.05 \left( \frac{r_a - r_ 0}{500 \km} \right)^{2/3}
\left( \frac{M_a}{0.1 M_\odot}\right)^{1/3}     \left( \frac{L_\nu}{5 \times 10^{52} \erg \s^{-1}} \right)^{-1/3}
\left( \frac{f}{0.1} \right)^{-1/3} \s.
\label{eq:ta}
\end{eqnarray}
{{{{ A similar acceleration time is estimated from numerical
results as we discussed in section \ref{sec:Time}, where this time
is marked $t_{\rm esc}$. }}}} Under these assumptions, the energy
of the ejected mass is
\begin{equation}
 E_a \simeq t_a f L_\nu \simeq
0.24 \left( \frac{r_a - r_ 0}{500 \km} \right)^{2/3}
\left( \frac{M_a}{0.1 M_\odot}\right)^{1/3}     \left( \frac{L_\nu}{5 \times 10^{52} \erg \s^{-1}} \right)^{2/3}
\left( \frac{f}{0.1} \right)^{2/3} \foe.
\label{eq:Ea}
\end{equation}
{{{{In these calculations, we assumed a constant neutrino
luminosity. As the  neutrino luminosity decreases with time
(e.g., \citealt{Fischeretal2012}), the term $fL_\nu$,
in equation (\ref{eq:Ea}) actually overestimates the available
energy. }}}} More typical values for acceleration over $\sim$$500
\km$ are $f < 0.1$ due to the low neutrino opacity (eq.
\ref{eq:depth}), and lower accelerated mass. These values give $
E_a <  0.2 ~ {\rm foe}$ as in Equation \ref{eq:Eshell}.

\section{THE ALMOST UNBOUND STALLED SHOCK}
\label{sec:shock}

The energy of the immediately post-shocked gas falling from thousands of km
to hundreds of km is close to zero before there is much neutrino cooling.
Whether the shocked gas falls or expands is a question of whether a small amount of
energy is added to revive the shock. When there are departures
from spherical symmetry, like the perturbations introduced by
\cite{CouchOtt2013} or instabilities in the post-shock region, in some areas
the extra energy comes at the expense of other areas. For example, a vortex
can add a positive velocity in the region of the flow where the flow goes
out. Even if the shock is revived, the energy limitations given in Sections
\ref{sec:Time} and \ref{sec:Energy} apply.
The SASI itself is a manifestation of the process where one region of the
stalled shock can go out in expense of other regions. The extra energy from
neutrino heating can even revive the entire sphere. However, the energy
gained by neutrino heating is limited.

A recent attempt to revive the stalled shock is that of \cite{CouchOtt2013},
who introduced perturbations to the Si/O layers, and found them to enable
shock revival under certain conditions.
What \cite{CouchOtt2013} term a successful explosion is actually a revival
of the stalled shock. They did not obtain the desired $\sim$$1 \foe$
explosion. As with many other simulations, small changes in the initial
conditions determine whether shock revival occurs or not. For example,
\cite{CouchOtt2013} find shock revival when their neutrino heat factor is
1.02, but not when it is 1.
They present their average shock position until it reaches a radius of 430
km at $t=0.32 \s$.
Examining their successful revival run presented in their figure 3, we find the
average shock outward velocity in the last part they show,
370 to 430 km, to be   $\langle v_{\rm shock} \rangle = 8000 \km \s^{-1}$. This is less
than 0.3 times the escape velocity at that radius. The shock does not seem to accelerate in
the last 50 km. Within $\Delta t \simeq 0.04 \s$ the shock will reach a
radius of about 700~km, where no more energy gain is possible \citep{Janka2001}. At $400 \km$ the neutrino optical depth is very small, $\tau
<0.1$. Indeed, at an average shock radius of $350 \km$ the heating
efficiency in their simulation $\eta$, defined as net heating rate divided
by $L_{\nu_e} + L_{\bar \nu_e}$, drops below 0.1. This implies that the gained
energy will be very small, $\Delta E < \tau L_\nu \Delta t < 0.2 \foe$.
We therefore estimate that even the perturbations introduced by
\cite{CouchOtt2013} will not bring the explosion, if occurs, close to $1
\foe$.

Let us quantify the statement of energy close-to-zero. We can make the following estimations based on the models of \citet{Woosleyetal2002} of massive stars prior to the collapse.
The gas at 2000 km has a specific gravitational energy of $e_{G0}=-10^{18} \erg \g^{-1}$ and a specific internal energy of $e_{I0}=5.5\times 10^{17} \erg \g^{-1}$.
After mass loss to neutrinos from the core, the inner mass reduces by $\sim$$10 \%$. However, by that time the shell that starts at
few$\times 1000 \km$ has been accelerated inward. So we take the total
specific energy to be as the pre-collapse energy.
As an example, we take the stalled shock to be at $r_s=200 \km$.
When reaching $r_s=200 \km$ the specific total energy $e_t=e_{I0}+e_{G0}$,
stays the same.
The specific gravitational energy is $e_{Gs} \simeq 10e_{G0} =-10^{19} \erg
\g^{-1}$, and the specific internal (thermal + kinetic+nuclear) energy, is
$e_{Is}=e_t-10e_{G0} \simeq 9.5\times 10^{18} \erg \g^{-1}$.
The net specific energy relative to gravitational energy in this
demonstrative example is
\begin{equation}
\xi_s \equiv \left| \frac {e_{Is}}{e_{Gs}}\right|\simeq 0.95.
\label{eq:Es}
\end{equation}
The mass is very close to be unbound. Small amount of net heating can revive
the shock. For a typical mass in the gain region of $M_g \la 0.05 M_\odot$ (e.g., \citealt{Couch2013ApJ}), an extra energy of $\Delta
E=5 \times 10^{49} \erg = 0.05 \foe$ will revive the shock.

\section{ENERGY AVAILABLE FROM RECOMBINATION}
\label{sec:recombination}

Adding nuclear energy of free nucleons does not change the above property of an almost unbound stalled shock, and the conclusion of low `explosion' energy.
Consider the scenario where disintegration of nuclei form free nucleons beyond the stalled shock, and the available nuclear energy is reused later
after the free nucleons are accelerated outwards by neutrinos \citep{Jankaetal2012}.
When the nucleons recombine to form heavy nuclei an energy of up to $9~ \rm{MeV}$ per nucleon can in principle be used to
explode the star \citep{Jankaetal2012}. A mass of $0.06M_\odot$ in the gain region can then release in principle $\sim 10^{51} \erg$ \citep{Scheck2006}.

However, the recombination of free nucleons to alpha particles, a process that uses $7 \MeV$ from the $9 \MeV$ available in forming silicon, starts when the reviving post-shock gas
reaches $r \sim 250 \km$ \citep{Fernandez2009}.
The energy released by recombination accelerates the material \citep{Fernandez2009},
which results in a shorter acceleration time than given in Equation (\ref{eq:ta}).
This further lowers the energy that can be supplied by neutrinos below that given in Equations (\ref{eq:Eshell}) and (\ref{eq:Ea}).

The energy available from recombination is limited as well. From
Figure 5 of \cite{Fernandez2009}, we find the total fraction of
$\alpha$ particles in the gas inside the shock when the shock
radius is $500 \km$ to be $\rm {X}_\alpha \la 0.5$; the fraction
just behind the shock front at $500 \km$ is $\rm {X}_\alpha \simeq
0.9$. In the results of \cite{Fernandez2009} the fraction of
$\alpha$ particles increases as the shock moves outward. For this
fraction, the average energy available from recombination is ~ 5
MeV/ nucleaon \citep{Jankaetal2012}. However, the shock is only at
$500 \km$ and a large fraction of the mass is much deeper. As the
shock expands further, the fraction of $\alpha$ particles will
increase and the available energy will decrease. Taking a mass in
the gain region of $M_{\rm gain} \la 0.05 M_\odot$ (e.g.,
\citealt{Couch2013ApJ}), we find the `explosion' energy to be
$E_{\rm nuc} \la 0.5 \foe$. In some 2D simulations the mass in the
gain layer is $\ga 0.05 M_\odot$, but in 3D simulations the gain
layer has lower mass than in 2D simulations (e.g.,
\citealt{Couch2013}). Over all, the available energy without
neutrino winds or jets is $< 0.5 \foe$. This value is an upper
limit and consistent with many of the simulations summarized in
section \ref{sec:zintro} that achieve much lower energies or do
not revive the shock at all.

It should be emphasized that the recombination is not a new energy source, as the thermal energy of the shocked gas is used to disintegrate the nuclei.
The recombination is the re-usage of this energy. The extra energy must come from neutrinos that lift the free nucleons to larger radii.
The total available energy from recombination is proportional to the mass of the free nucleons that are lifted from small $r \la 150 \km$ to large radii $r \ga 500 \km$.
However, the amount of mass that can be accumulated at small radii is limited because if the density is too high then cooling overcomes neutrino heating,
and the shock will not be revive.

\cite{Yamamotoetal2013} preformed 1D and 2D simulations of shock revival and
examined explosion energy including recombination and shock nuclear burning.
They tuned the neutrino luminosity to a critical value that gives successful
explosions. Their successful runs have shock relaunch times of $0.3-0.4 \s$
in 2D flows. The explosion energy in these runs is in the range of $0.6-1.5
\foe$.
We note the following regarding their tuned calculations:
\newline
(1) \cite{Yamamotoetal2013} assume that neutrino heating alone revives the
stalled shock. Then they can use the entire recombination energy to explode
the rest of the star. The more realistic calculations of
\cite{Fernandez2009} show that at least half the recombination energy is
required to help revive the shock.
\newline
(2) The above assumption implies the need for high neutrino luminosity.
Indeed, in \cite{Yamamotoetal2013}  successful 2D runs the required critical
neutrino luminosities are  $L_{\nu,c}= L_{\bar\nu,c}=4.8 \times 10^{52} \erg
\s^{-1}$ and $4.5 \times 10^{52} \erg \s^{-1}$ for shock relaunching times
of $0.3 \s$ and $0.4 \s$, respectively.
These neutrino luminosities are $\sim 50\%$ higher than what most realistic numerical simulations find, e.g., \cite{Fischeretal2012},
and $\sim 30 \%$ higher than the neutrino luminosities obtained by \cite{Mueller2012} who included general relativistic effects.
Interestingly, \cite{Mueller2012} find for their $11 M_\odot$ model that recombination of nucleons and $\alpha$-particles in the ejecta
would provide an additional energy of $E_{\rm rec} \simeq 0.02 \foe$. For their $15 M_\odot$ model they argue that
burning in the shock will add on the order of $0.1-0.2 \foe$ or more.
\newline
(3) The contribution of nuclear and recombination energies to the diagnostic explosion energy of \cite{Yamamotoetal2013} are very similar to
the contribution of neutrino heating.

Based on these points we can use a more realistic value of neutrino heating,
$E_\nu < 0.2 \foe$, and conclude that the combined explosion energy in realistic simulations will be $E_{\rm exp} <0.5 \foe$. Again we reach the conclusion that including
recombination energy will at most bring the explosion energy to $E_{\rm exp} <0.5 \foe$.
Although close to the canonical $1 \foe$ value, one must keep in mind that this value is obtained with very favorable conditions, and in scaled, rather than realistic, simulations.
In more realistic simulations the recombination energy is found to be $E_{\rm rec} \la 0.2 \foe$, e.g., \cite{Mueller2012}.

\section{DISCUSSION AND SUMMARY}
\label{sec:summary}
Using simple estimates of a spherically-symmetric mass ejection by
neutrino flux in core-collapse supernovae (CCSNe), we found that
in the delayed-neutrino mechanism \citep{bethe1985}, {where the
main energy source of the} explosion is due to neutrino heating in
the gain region, the explosion energy is limited to $ E \la 0.5
\foe$, with a more likely limit of $0.3 \foe$ (eq. \ref{eq:Eshell}
and \ref{eq:Ea}). This falls short of what is required in most
CCSNe.

Although our simple analytical estimates are limited to
spherically symmetric outflows, they none-the-less catch the
essence of the delayed-neutrino mechanism. In a non-spherical
flow, instabilities, such as neutrino-driven convection and the
standing accretion shock instability (SASI), play a major role
(e.g., \citealt{Hanke2013}). Such instabilities allow inflow and
outflow to occur simultaneously. Still, recent and highly
sophisticated 3D simulations with enough details to resolve such
instabilities do not obtain enough energy to revive the stalled
shock, (e.g., \citealt{Janka2013}). The energy that can be used from
the neutrino flux might, under favorite conditions, revive the
stalled shock, but cannot lead to explosions with energies of $E_e
\ga 0.3 \foe$.

Our conclusion holds as long as no substantially new
ingredient is added to the delayed-neutrino mechanism. Such an
ingredient can be a strong wind, as was applied by artificial
energy deposition by \cite{2010PhRvD..82j3016N,Nordhaus2012}. In
their 2.5D simulations, \cite{Scheck2006} achieved explosion that
was mainly driven by a continuous wind. The problem we see with
winds is that they are less efficient than jets. Indeed, in order
to obtain an explosion the winds in the simulations of
\cite{Scheck2006} had to be massive. For that, in cases where they
obtained energetic enough explosions the final mass of the NS was
low  $(M_{\rm NS} <1.3 M_\odot)$.  Such a wind must be active
while accretion takes place; the accretion is required to supply
the energy \citep{Marek2009}.

{With the severe problems encountered by neutrino heating,
research groups have turned to study dynamical processes.
\cite{CouchOtt2013}, \cite{CouchOtt2014}, and
\cite{MuellerJanka2014b} argued that the effective turbulent ram
pressure exerted on the stalled shock allows shock revival with
less neutrino heating than 1D models. However,
\cite{Abdikamalovetal2014} found that increasing the numerical
resolution allows cascade of turbulent energy to smaller scales,
and the shock revival becomes harder to achieve at high numerical
resolutions.}

{Another dynamical process is a collimated wind blown by the newly
formed NS. \citet{Bruennetal2015} performed 2D simulations up to
$1.4 \s$ post-bounce, and obtained an explosion energy of $0.3-0.9
\foe$, depending on the stellar model (initial mass without
rotation). They find the main energy source to be by what they
term an `enthalpy flux'. This is actually a wind, mainly along the
imposed symmetry axis, i.e., a collimated wind. This wind is
driven by the inflowing (accreted) gas. At some instant, their
results show jet-like outflows along the symmetry axis. It seems
that the collimated wind is induced by the numerical grid.
Contrast that to their corresponding 3D simulations
\citep{Mezzacappaetal2014} } which show no such explosion.
\cite{Mezzacappaetal2014} present one new result of a 3D run for
their $15 M_\odot$ model at $t=267 \ms$ post-bounce. We estimate
the average shock radius at that time to be $\sim 220 \km$. This
is very similar to their 1D results \citep{Bruenn2013}, where the
shock radius is much smaller than in their 2D simulations, and
where no explosions occur.  {Non-the-less, the results of
\cite{Bruennetal2015} show the great potential of an
inflow-outflow mechanism in exploding CCSNe. An inflow-outflow
situation with collimated outflows over a relatively long time
naturally occurs with jets launched by accretion disks, without
the numerically induced symmetry axis in 2D grids.}

For the above,  the lack of persisting success, and possibly
failure, of the delayed-neutrino mechanism calls for a paradigm
shift. As well, the rich variety of CCSN properties (e.g.,
\citealt{Arcavi2012}) further emphasizes the need to study
alternative models for CCSN explosions, some of which are based on
jet-driven explosions \citep{Janka2012}. In CCSNe simulations jets
have been shown to be launched when the pre-collapsing core posses
both a rapid rotation and a very strong magnetic field (e.g.
\citealt{LeBlanc1970, Meier1976, Bisnovatyi1976, Khokhlov1999,
MacFadyen2001,Hoflich2001, Woosley2005, Burrows2007,
Couch2009,Couch2011,Lazzati2011, TakiwakiKotake2011}). However,
these jets do not explode the core via a feedback mechanism, such
that they too often give extreme cases as gamma ray bursts, or
they fail to explode the star, e.g., \cite{mostaetal2014}. Recent
observations (e.g. \citealt{Milisavljevic2013,Lopez2013}) suggest
that jets might play a role in at least some CCSNe. Another
motivation to consider jet-driven explosion mechanisms is that
jets might supply the site for the r-process
\citep{Winteler2012,Papish2012b}. The question is whether the
accreted mass possesses sufficient specific angular momentum to
form an accretion disk. Persistent accretion disk requires the
pre-collapsing core to rotate fast, as in the magnetohydrodynamics
class of models ( e.g. \citealt{LeBlanc1970, Meier1976,
Bisnovatyi1976, Khokhlov1999, MacFadyen2001,Hoflich2001,
Woosley2005, Burrows2007, Couch2009,Couch2011,Lazzati2011}). Most
massive stars reach the core-collapse phase with a too slow core
rotation for the magnetorotational mechanism to be significant.

One alternative to the delayed-neutrino mechanism which overcomes
the angular momentum barrier is the so-called ``jittering-jet"
mechanism of \cite{Papish2011}. The jittering-jet mechanism
overcomes the requirement for rapid core rotation, and was
introduced as a mechanism to explode all CCSNe \citep{Papish2011,
Papish2012b, Papish2014}. The angular momentum source is the
convective regions in the core \citep{GilkisSoker2014}, and/or
instabilities in the shocked region of the collapsing core.
\cite{BlondinMezzacappa2007}, \cite{Fernandez2010}, and
\cite{Rantsiouetal2011} suggested that the source of the angular
momentum of pulsars is the spiral mode of the SASI.  In the
jittering-jet mechanism there is no need to revive the accretion
shock, and it is a mechanism based on a negative feedback cycle.
As long as the core was not exploded, the accretion continues.
After an energy equals several times the core binding energy is
deposited to the core by the jets, the star explodes.  This energy
amounts to $\sim 1 \foe$. If the feedback is less efficient, more
accretion is required to accumulate the required energy.  If the
efficiency is very low, the accreted mass onto the NS brings it to
collapse to a black hole and launch relativistic jets.  Namely, in
general, \emph{the less efficient the feedback mechanism is, the
more violent the explosion is} \citep{GilkisSoker2014}.

{{{{{ We thank the anonymous referee for helpful comments.
}}}}}
 This research was supported by the Asher Fund for Space Research
at the Technion, and a generous grant from the president of the
Technion Prof. Peretz Lavie.  OP is supported by the Gutwirth
Fellowship. JN is supported by an NSF award AST-1102738 and by
NASA HST grant AR-12146.01-A.



\label{lastpage}


\begin{thebibliography}{}\addcontentsline{toc}{section}{References}

\bibitem[Abdikamalov et al.(2014)]{Abdikamalovetal2014} Abdikamalov, E.,
Ott, C.~D., Radice, D., et al.\ 2014, arXiv:1409.7078

\bibitem[Arcavi et al.(2012)]{Arcavi2012} Arcavi, I., Gal-Yam, A., Cenko, S.~B., et al.\ 2012, \apjl, 756, L30

\bibitem[Baade \& Zwicky(1934)]{Badde1934} Baade, W., \& Zwicky, F.\ 1934, Physical Review, 46, 76

\bibitem[Bethe \& Wilson(1985)]{bethe1985} Bethe, H.~A., \& Wilson, J.~R.\ 1985, \apj, 295, 14

\bibitem[Bisnovatyi-Kogan et al.(1976)]{Bisnovatyi1976} Bisnovatyi-Kogan, G.~S., Popov, I.~P., \& Samokhin, A.~A.\ 1976, \apss, 41, 287

\bibitem[Blondin et al.(2003)]{BlondinMezzacappa2003} Blondin, J.~M., Mezzacappa, A., \& DeMarino, C.\ 2003, \apj, 584, 971

\bibitem[Blondin \& Mezzacappa(2007)]{BlondinMezzacappa2007} Blondin, J.~M., \& Mezzacappa, A.\ 2007, \nat, 445, 58

\bibitem[Brandt et al.(2011)]{Brandt2011} Brandt, T.~D., Burrows, A., Ott, C.~D., \& Livne, E.\ 2011, ApJ, 728, 8

\bibitem[Bruenn et al.(2014)]{Bruennetal2015} Bruenn, S.~W., Lentz,
E.~J., Hix, W.~R., et al.\ 2014, arXiv:1409.5779

\bibitem[Bruenn et al.(2013)]{Bruenn2013} Bruenn, S.~W., Mezzacappa, A., Hix, W.~R., et al.\ 2013, \apjl, 767, L6

\bibitem[Buras et al.(2003)]{Buras2003} Buras, R., Rampp, M., Janka, H.-T., \& Kifonidis, K.\ 2003, Physical Review Letters, 90, 241101

\bibitem[Buras et al.(2006a)]{Buras2006a} Buras, R., Rampp, M., Janka, H.-T., \& Kifonidis, K.\ 2006, \aap, 447, 1049

\bibitem[Buras et al.(2006b)]{Buras2006b} Buras, R., Janka, H.-T., Rampp, M., \& Kifonidis, K.\ 2006, \aap, 457, 281

\bibitem[Burrows \& Lattimer(1985)]{Burrows1985} Burrows, A., \& Lattimer, J.~M.\ 1985, \apjl, 299, L19

\bibitem[Burrows \& Goshy(1993)]{Burrows1993} Burrows, A., \& Goshy, J.\ 1993, \apjl, 416, L75

\bibitem[Burrows et al.(1995)]{Burrows1995} Burrows, A., Hayes, J., \& Fryxell, B.~A.\ 1995, \apj, 450, 830

\bibitem[Burrows et al.(2007)]{Burrows2007} Burrows, A., Dessart,
L., Livne, E., Ott, C.~D., \& Murphy, J.\ 2007, \apj, 664, 416

\bibitem[Colgate \& White(1966)]{Colgate1966} Colgate, S.~A., \& White, R.~H.\ 1966, \apj, 143, 626

\bibitem[Couch (2013)]{Couch2013} Couch, S.~M. 2013, Presented in the Fifty-one erg meeting, Raleigh, May 2013.

\bibitem[Couch(2013)]{Couch2013ApJ} Couch, S.~M.\ 2013, \apj, 775,
35

\bibitem[Couch \& Ott(2013)]{CouchOtt2013} Couch, S.~M., \& Ott, C.~D.\ 2013, \apjl, 778, L7

{{{ {\bibitem[Couch \& Ott(2014)]{CouchOtt2014} Couch, S.~M., \&
Ott, C.~D.\ 2014, arXiv:1408.1399} }}}


\bibitem[Couch 
\& O'Connor(2014)]{couch2013arXiv} Couch, S.~M., \& O'Connor, E.~P.\ 2014, \apj, 785, 123 



\bibitem[Couch et al.(2009)]{Couch2009} Couch, S.~M., Wheeler, J.~C., \& Milosavljevi{\'c}, M.\ 2009, \apj, 696, 953

\bibitem[Couch et al.(2011)]{Couch2011} Couch, S.~M., Pooley, D., Wheeler, J.~C., \& Milosavljevi{\'c}, M.\ 2011, \apj, 727, 104

\bibitem[Dolence et al.(2013)]{Dolence2013} Dolence, J.~C., Burrows, A., Murphy, J.~W., \& Nordhaus, J.\ 2013, \apj, 765, 110

\bibitem[Dolence et al.(2014)]{Dolenceetal2014} Dolence, J.~C.,
Burrows, A., \& Zhang, W.\ 2014, arXiv:1403.6115

\bibitem[Fern{\'a}ndez \& Thompson(2009)]{Fernandez2009} Fern{\'a}ndez, R., \& Thompson, C.\ 2009, \apj, 703, 1464

\bibitem[Fern{\'a}ndez(2010)]{Fernandez2010} Fern{\'a}ndez, R.\ 2010, \apj, 725, 1563

\bibitem[Fischer et al.(2012)]{Fischeretal2012} Fischer, T.,
Mart{\'{\i}}nez-Pinedo, G., Hempel, M.,
\& Liebend{\"o}rfer, M.\ 2012, \prd, 85, 083003

\bibitem[Fryer \& Warren(2002)]{Fryer2002} Fryer, C.~L., \& Warren, M.~S.\ 2002, \apjl, 574, L65

\bibitem[Gilkis 
\& Soker(2014)]{GilkisSoker2014} Gilkis, A., \& Soker, N.\ 2014, \mnras, 439, 4011 


\bibitem[Hanke et al.(2012)]{Hanke2012} Hanke, F., Marek, A., M{\"u}ller, B., \& Janka, H.-T.\ 2012, \apj, 755, 138

\bibitem[Hanke et al.(2013)]{Hanke2013} Hanke, F., Mueller, B., Wongwathanarat, A., Marek, A., \& Janka, H.-T.\ 2013, \apj, 770, 66

\bibitem[H{\"o}flich et al.(2001)]{Hoflich2001} H{\"o}flich, P., Khokhlov, A., \& Wang, L.\ 2001, 20th Texas Symposium on relativistic astrophysics, 586, 459


\bibitem[Janka \& Mueller(1996)]{Janka1996} Janka, H.-T., \& Mueller, E.\ 1996, \aap, 306, 167

\bibitem[Janka(2001)]{Janka2001} Janka, H.-T.\ 2001, A\&A, 368, 527

\bibitem[Janka(2012)]{Janka2012} Janka, H.-T.\ 2012, Annual Review of Nuclear and Particle Science, 62, 407

\bibitem[Janka(2013)]{Janka2013} Janka, H.-T.\ 2013, Presented in the Fifty-one erg meeting, Raleigh, May 2013.

\bibitem[Janka et al.(2012)]{Jankaetal2012} Janka, H.-T., Hanke, F., H{\"u}depohl, L., et al.\ 2012, Progress of Theoretical and Experimental Physics, 2012, 010000

\bibitem[Khokhlov et al.(1999)]{Khokhlov1999} Khokhlov, A.~M., H{\"o}flich, P.~A., Oran, E.~S., et al.\ 1999, \apjl, 524, L107

\bibitem[Kitaura et al.(2006)]{Kitaura2006} Kitaura, F.~S., Janka, H.-T., \& Hillebrandt, W.\ 2006, \aap, 450, 345

\bibitem[Kuroda et al.(2012)]{Kuroda2012} Kuroda, T., Kotake, K., \& Takiwaki, T.\ 2012, \apj, 755, 11

\bibitem[Lazzati et al.(2012)]{Lazzati2011} Lazzati, D., Morsony, 
B.~J., Blackwell, C.~H., \& Begelman, M.~C.\ 2012, \apj, 750, 68 

\bibitem[LeBlanc \& Wilson(1970)]{LeBlanc1970} LeBlanc, J.~M., \& Wilson, J.~R.\ 1970, \apj, 161, 541

\bibitem[Liebend{\"o}rfer et al.(2001)]{Liebendorferetal} Liebend{\"o}rfer, M., Mezzacappa, A., Thielemann, F.-K., et al.\ 2001, \prd, 63, 103004

\bibitem[Liebend{\"o}rfer et al.(2005)]{Liebend2005} Liebend{\"o}rfer, M., Rampp, M., Janka, H.-T., \& Mezzacappa, A.\ 2005, \apj, 620, 840

\bibitem[Lopez et al.(2013)]{Lopez2013} Lopez, L.~A., Ramirez-Ruiz, E., Castro, D., \& Pearson, S.\ 2013, \apj, 764, 50

\bibitem[MacFadyen et al.(2001)]{MacFadyen2001} MacFadyen, A.~I., Woosley, S.~E., \& Heger, A.\ 2001, \apj, 550, 410

\bibitem[Marek \& Janka(2009)]{Marek2009} Marek, A., \& Janka, H.-T.\ 2009, \apj, 694, 664

\bibitem[Meier et al.(1976)]{Meier1976} Meier, D.~L., Epstein, R.~I., Arnett, W.~D., \& Schramm, D.~N.\ 1976, \apj, 204, 869

\bibitem[Mezzacappa et al.(2001)]{Mezzacappa2001} Mezzacappa, A., Liebend{\"o}rfer, M., Messer, O.~E., et al.\ 2001, Physical Review Letters, 86, 1935

\bibitem[Mezzacappa et al.(2014)]{Mezzacappaetal2014} Mezzacappa, A., 
Bruenn, S.~W., Lentz, E.~J., et al.\ 2014, 8th International Conference of 
Numerical Modeling of Space Plasma Flows (ASTRONUM 2013), 488, 102 

\bibitem[Milisavljevic et al.(2013)]{Milisavljevic2013} Milisavljevic, 
D., Soderberg, A.~M., Margutti, R., et al.\ 2013, \apjl, 770, LL38 

\bibitem[M{\"o}sta et al.(2014)]{mostaetal2014} M{\"o}sta, P., 
Richers, S., Ott, C.~D., et al.\ 2014, \apjl, 785, LL29 

\bibitem[Mueller et al.(2012)]{Mueller2012} Mueller, B., Janka, H.-T., \& Marek, A.\ 2012, \apj, 756, 84


\bibitem[M{\"u}ller 
\& Janka(2014)]{MuellerJanka2014} M{\"u}ller, B., \& Janka, H.-T.\ 2014, \apj, 788, 82 

\bibitem[Mueller \& Janka(2014)]{MuellerJanka2014b} Mueller, B.,
\& Janka, H.-T.\ 2014, arXiv:1409.4783

\bibitem[Nakamura et al.(2014)]{Nakamuraetal2014} Nakamura, K., Kuroda, 
T., Takiwaki, T., \& Kotake, K.\ 2014, \apj, 793, 45 

\bibitem[Nomoto \& Hashimoto(1988)]{Nomoto1988} Nomoto, K., \& Hashimoto, M.\ 1988, \physrep, 163, 13

\bibitem[Nordhaus et al.(2010)]{2010PhRvD..82j3016N} Nordhaus, J., Brandt, T.~D., Burrows, A., Livne, E., \& Ott, C.~D.\ 2010, \prd, 82, 103016

\bibitem[Nordhaus et al.(2010)]{Nordhaus2010} Nordhaus, J., Burrows, A., Almgren, A., \& Bell, J.\ 2010, \apj, 720, 694

\bibitem[Nordhaus et al.(2012)]{Nordhaus2012} Nordhaus, J., Brandt, T., Burrows, A., \& Almgren, A.\ 2011, \mnras, 423, 1805


\bibitem[Ott et al.(2008)]{Ott2008} Ott, C.~D., Burrows, A., Dessart, L., \& Livne, E.\ 2008, \apj, 685, 1069

\bibitem[Papish \& Soker(2011)]{Papish2011} Papish, O., \& Soker, N.\ 2011, \mnras, 416, 1697

\bibitem[Papish \& Soker(2012a)]{Papish2012a} Papish, O., \& Soker, N.\ 2012a, Death of Massive Stars: Supernovae and Gamma-Ray Bursts, 279, 377

\bibitem[Papish \& Soker(2012b)]{Papish2012b} Papish, O., \& Soker, N.\ 2012b, \mnras, 421, 2763

\bibitem[Papish \& Soker(2014)]{Papish2014} Papish, O., \& Soker, N.\ 2014, \mnras,


\bibitem[Rampp \& Janka(2000)]{Rampp2000} Rampp, M., \& Janka, H.-T.\ 2000, \apjl, 539, L33

\bibitem[Rantsiou et al.(2011)]{Rantsiouetal2011} {{{ {Rantsiou, E., Burrows, A., Nordhaus, J., \& Almgren, A.\ 2011, \apj, 732, 57}  }}}

\bibitem[Roy et al.(2013)]{Roy2013} Roy, R., Kumar, B., Maund, 
J.~R., et al.\ 2013, \mnras, 434, 2032 

\bibitem[Scheck et al.(2006)]{Scheck2006} Scheck, L., Kifonidis, K., Janka, H.-T., Mueller, E.\ 2006, A\&A, 457, 963

\bibitem[Suwa(2014)]{Suwa2013} Suwa, Y.\ 2014, \pasj, 66, LL1 

\bibitem[Suwa et al.(2010)]{Suwaetal2010} Suwa, Y., Kotake, K., Takiwaki, T., et al.\ 2010, \pasj, 62, L49


\bibitem[Sawai 
\& Yamada(2014)]{SawaiYamada2014} Sawai, H., \& Yamada, S.\ 2014, \apjl, 784, LL10 


{{{ \bibitem[Takiwaki
\& Kotake(2011)]{TakiwakiKotake2011} Takiwaki, T., \& Kotake, K.\ 2011, \apj, 743, 30 }}}

\bibitem[Takiwaki et al.(2014)]{Takiwakietal2013} Takiwaki, T., Kotake,
K., \& Suwa, Y.\ 2014, \apj, 786, 83

\bibitem[Takaki et al.(2013)]{Takaki2013} Takaki, K., Kawabata, 
K.~S., Yamanaka, M., et al.\ 2013, \apjl, 772, LL17 

\bibitem[Ugliano et al.(2012)]{Ugliano2012} Ugliano, M., Janka, H.-T., Marek, A., \& Arcones, A.\ 2012, \apj, 757, 69

\bibitem[Wilson(1985)]{Wilson1985} Wilson, J.~R.\ 1985, Numerical Astrophysics, 422

\bibitem[Winteler et al.(2012)]{Winteler2012} Winteler, C., K{\"a}ppeli, R., Perego, A., et al.\ 2012, \apjl, 750, L22

\bibitem[Woosley \& Janka(2005)]{Woosley2005} Woosley, S., \& Janka, T.\ 2005, Nature Physics, 1, 147

\bibitem[Woosley et al.(2002)]{Woosleyetal2002} Woosley, S.~E., Heger,
A., \& Weaver, T.~A.\ 2002, Reviews of Modern Physics, 74, 1015

\bibitem[Yamamoto et al.(2013)]{Yamamotoetal2013} Yamamoto, Y.,  Fujimoto, S.-i., Nagakura, H., \& Yamada, S.\ 2013, \apj, 771, 27

\bibitem[Zhang et al.(2013)]{Zhang2013} Zhang, W., Howell, L., Almgren, A., et al.\ 2013, \apjs, 204, 7

\end{thebibliography}
\end{document}